\documentclass{aastex62}

\received{March 15, 2019}
\revised{April 9, 2019}
\accepted{May 21, 2019}

\shorttitle{Stochastic heating radial evolution}
\shortauthors{Martinovi\'c \& Others}

\begin{document}

\title{Radial evolution of stochastic heating in low-$\beta$ solar wind}

\correspondingauthor{M. M. Martinovi\'c}
\email{mmartinovic@email.arizona.edu}

\author[0000-0002-7365-0472]{Mihailo M. Martinovi\'c}
\affiliation{Lunar and Planetary Laboratory, University of Arizona, Tucson, AZ 85719, USA.}
\affiliation{LESIA, Observatoire de Paris, Meudon, France.}
\affiliation{Department of Astronomy, Faculty of Mathematics, University of Belgrade, Serbia.}

\author[0000-0001-6038-1923]{Kristopher G. Klein}
\affiliation{Lunar and Planetary Laboratory, University of Arizona, Tucson, AZ 85719, USA.}

\author[0000-0002-2358-6628]{Sofiane Bourouaine}
\affiliation{Florida Institute of Technology, Melbourne, FL, United States}

\begin{abstract}

We consider the radial evolution of perpendicular ion heating due to the violation of magnetic moment conservation caused by Alfv{\' e}n and kinetic Alfv{\' e}n wave turbulence. This process, referred to as stochastic heating, can be quantified by the ratio between the average velocity fluctuations at the ion gyroradius and the perpendicular ion thermal speed $\epsilon \equiv \delta v / v_{t\perp}$. Using 17 years of Helios observations, we constrain how much energy could be dissipated by this mechanism between 0.29 and 0.98 au. We find that stochastic heating likely operates throughout the entire inner heliosphere, but that its radial dependence is steeper than that of empirically derived dissipation rates, being $r^{-2.5}$ compared with $r^{-2.08}$. This difference is significantly increased in fast solar wind streams to $r^{-3.1}$ compared with $r^{-1.8}$. 

\end{abstract}


\section{Introduction}
\label{sec:Intro}

Over the last six decades, numerous hydrodynamic models of the solar wind \citep{Parker_1958,Sturrock_1966,Wolff_1971}, which rely on thermal conduction as the driver of the solar wind heating and acceleration, were developed. These models are unable to explain the existence of fast solar wind streams and large proton temperatures measured at 1 au (see e.g. \citet{Echim_2011} and references therein), suggesting that there are additional heating mechanisms operating above the coronal base. Among other candidate processes \citep{Markovskii_2010,Drake_2012,Osman_2012,Kasper_2013,Vech_2018b}, Alfv{\' e}n wave (AW) turbulent dissipation has been widely examined \citep{Coleman_1968,Hollweg_1973,Villante_1980,Velli_1989,Matthaeus_1999} as one of the possible answers to this open question, ever since large amplitude AWs were first observed in the interplanetary medium \citep{Belcher_1969,Belcher_1971}.  

In this work we focus on heating of the solar wind heating due to turbulent Alfv\'enic fluctuations with perpendicular scales of order the proton gyroradius ($k_\perp \rho \sim 1$, where $k_\perp$ is the component of the wave vector $\mathbf{k}$ perpendicular to the magnetic field $\mathbf{B}$). At these scales, the turbulent Alfv\'enic cascade transforms into a kinetic Alfv{\' e}n wave (KAW) cascade (see e.g. \citet{Bale_2005,Howes_2008b}). As the turbulent power is wavevector anisotropic with $k_\perp \gg k_\parallel$ \citep{Goldreich_1995,Cho_2004,Schekochihin_2009,Chen_2010,Mallet_2015}, the KAWs have frequencies smaller than the proton cyclotron frequency $\Omega_p$ \citep{Howes_2008a}, making the collisionless damping due to cyclotron resonances negligible \citep{Lehe_2009}. Also, for the magnetically dominated, low-$\beta$ environment expected close to the Sun, the ion thermal speed $v_t$ is much smaller than the Alfv{\' e}n speed $v_A$, quenching Landau or transit time damping, which requires $\omega \approx k_{\parallel}v_{t\parallel}$ \citep{Quataert_1998,Hollweg_1999}, where $v_{t||}$ is the parallel thermal speed.While dissipation via Landau damping has been identified in magnetosheath turbulence \citep{Chen_2019} and gyrokinetic numerical simulations of turbulence \citep{Klein_2017_JPlPh} using the field-particle correlation technique \citep{Klein_2016,Howes_2017_JPP}, such linear mechanisms are not accessible for the measured particle and wave parameters at proton gyroradius scales in the inner Heliosphere. This difficulty motivated numerous authors \citep{Hollweg_2002,Dmitruk_2002,Dmitruk_2004,Drake_2009,Lehe_2009} to investigate perpendicular ion heating due to non-linear processes. 

Here, we examine the effects of stochastic heating (SH) by low frequency turbulent Alfv\'enic fluctuations near the proton gyroradius scale. This mechanism is based on the idea that magnetic moment $\mu_m = m_p v_\perp^2/2 B$ invariance, $m_p$ being the proton mass, is violated in a turbulent plasma if the amplitude of the velocity fluctuations is sufficiently large, leading to heating of the protons. In the theoretical work of \citet{Chandran_2010}, it was shown that this process can be phenomenologically quantified by the parameter

\begin{equation}
\epsilon = \frac{\delta v}{v_{t\perp}}
\label{eq:Epsilon}
\end{equation}

\noindent where $\delta v$ is the rms amplitude of velocity fluctuations at the proton gyroradius scale, $v_{t\perp} = \sqrt{{2 k_b T_{\perp}}/{m_p}}$ is the perpendicular thermal speed, and $k_b$ is the Boltzmann constant. As $\epsilon$ increases, the particle orbits become increasingly chaotic and depart from their smooth gyromotion, leading to perpendicular diffusion. 
Given a particular amplitude of turbulent fluctuations $\delta v$, and assuming $\beta_{||} \lesssim 1$, the associated heating rate due to SH is given by
\begin{equation}
Q_\perp = \frac{c_1 (\delta v)^3}{\rho_p} \exp \left[-\frac{c_2}{\epsilon}\right],
\label{eq:Heating_Rate}
\end{equation}
where $c_1$ and $c_2$ are dimensionless parameters of order unity that can be extracted from numerical simulations.
\citet{Chandran_2010} found that for test particle simulations with $\epsilon_{\rm crit} = 0.19$, $Q_\perp$ equalled half of the turbulent cascade power at $k_\perp \rho \sim 1$. 

Observational tests for the presence of SH were performed by \citet{Bourouaine_2013}, who calculated $\epsilon$ for low-$\beta$, fast solar wind using Helios data at three different radial distances (0.29, 0.4 and $0.64$ au). At all three distances, similar values of $\epsilon \approx 0.048$ were found. The authors concluded that, for values of $c_1$ and $c_2$ consistent with RMHD turbulence simulations, $Q_\perp$ is comparable to an empirically derived heating rate. A more recent statistical study performed by \citet{Vech_2017} using Wind data at 1 au reported that the proton temperature anisotropy $T_\perp / T_{||}$, likely the result of some preferential perpendicular heating mechanism such as SH, and the scalar proton temperature $T_p$ linearly increased with $\epsilon \geq \epsilon_{\rm{crit}} \approx 0.025$, a turbulence amplitude for which SH is expected to be negligible. These results present a dilemma; does SH continuously operate throughout the entire inner heliosphere, or are observations at $1$ au a remnant of processes that happened days ago, when the solar wind was closer to the Sun?

Pursuing an answer to this question, we use a method analogous to the one described in \citet{Bourouaine_2013} to process 17 years of observations from both Helios 1 and 2 \citep{Porsche_1981a,Porsche_1981b} and investigate possible dependencies of $\epsilon$ and $Q_\perp$ as a function of the radial distance from the Sun. An overview of the method is given in Section \ref{sec:Method}. Processing of the E2 magnetometer data required removing the effects of instrument noise and saturation, as well as the spacecraft spin, as explained in Section \ref{sec:Instrumentation}. Section \ref{sec:Results} contains the results of the statistical study, which we compare with theoretical \citep{Chandran_2011} and empirical \citep{Hellinger_2011,Hellinger_2013} models in order to examine the importance of SH at different radial distances, as well as inferences for enhanced heating at distances closer than the Helios perihelia as predicted by \citet{Kasper_2017}. Discussion and possible implications of our results, as well as their limitations, are discussed in Section \ref{sec:Discussion}.

\section{Method}
\label{sec:Method}

\begin{figure*}
\gridline{\fig{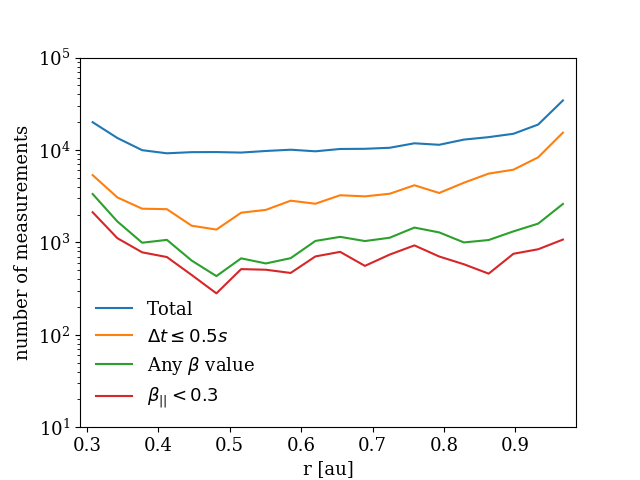}{0.49\textwidth}{(a)}
          \fig{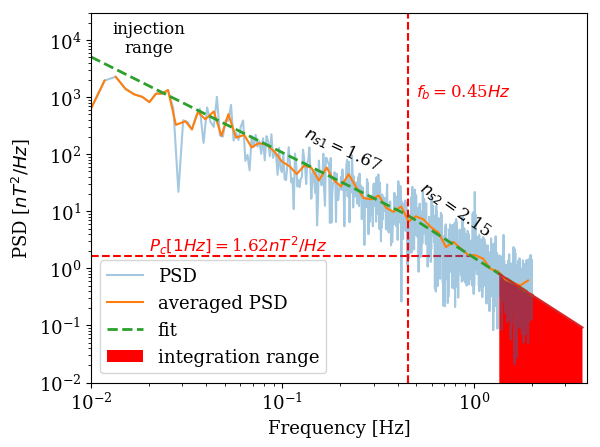}{0.49\textwidth}{(b)}
          }
    \caption{(a) Number of 10-minute intervals processed in this study. (b) Example of a processed 10-minute power spectrum, observed by Helios 2 between 03:50 and 04:00 on April 14, 1976.}
    \label{fig:Spectrum}
\end{figure*}

Our method of data processing is based on that described in \citet{Bourouaine_2013}. We use the Helios E1 instrument \citep{Schwenn_1975} proton corefit data set with $\sim 40s$ time resolution provided by \citet{Stansby_2018} to obtain 10-minute averages of parallel and perpendicular proton temperatures, $T_\perp$ and $T_\parallel$ and mean angle $\theta$ between the solar wind flow and the magnetic field $\mathbf{B}$. The proton density $n_p$, density of $\alpha$ particles $n_{\alpha}$ and solar wind bulk velocity $v_{sw}$ were obtained from the data set on NASA CDAWeb \citep{CDAWeb_2018b}. In periods when data was missing from the CDAweb data set, we used corefit values. We note that the corefit data set only considers the core component of the proton distribution for densities, bulk velocities, and temperatures; proton density from the corefit data set is shown to be about $80\%$ of the total proton density, while $v_{sw}$ is on average $\sim 1.5\%$ lower. For these intervals, we used an averaged value of $n_\alpha \approx 0.04 n_p$ \citep{Matthias_2001}. This approach enables us to work with $48\%$ more measurements (259801 instead of 175834, see Figure \ref{fig:Spectrum}(a)), while increasing the uncertainty in the results by up to $10\%$, less than the uncertainties that emerge from E2 instrument data processing and model parameters (Sections \ref{sec:Instrumentation} and \ref{sec:Results}).
High resolution magnetic field data from the E2 magnetometer \citep{Musmann_1975} is provided by the instrument team \citep{Cologne_2017_E2}. 
From this data set, we calculate the parallel plasma $\beta$, proton gyroradius and proton gyrofrequency, defined as

\begin{eqnarray}
\beta_{||} = \frac{2 \mu_0 n_p k_b T_{||}}{B^2} \label{eq:beta_approximation} \\
\rho_p = \frac{v_{t\perp}}{\Omega_p} \\
\Omega_p = \frac{e_c B}{m_p}
\end{eqnarray}

\noindent where constants $\mu_0$ and $e_c$ stand for the magnetic permeability of vacuum and the unity charge, respectively. 

Given that beam-like components of the proton velocity distribution appear mostly in the direction parallel to $\mathbf{B}$ \citep{Marsch_1982}, we assume the $T_\perp$ extracted from the corefit dataset is a good measure of the total perpendicular temperature. On the other hand, the scalar temperature reported by the corefit and core-and-beam data sets can vary significantly, with their ratio varying from 0.5 to 2, where the corefit temperature is $\sim 5\%$ lower on average \citep{Stansby_2018}. Consequently, our results for $T_\parallel$ may be slightly underestimated, lowering the accuracy of our low-$\beta_\parallel$ criteria given in Equation \ref{eq:beta_approximation}, but not to a degree that should affect the results presented in this work.

In order to derive values for $\epsilon$ and $Q_\perp$, we must determine the amplitude of the velocity fluctuations at the proton gyroscale, $\delta v$. As measurements of the proton velocity by Helios are not sufficiently fast to capture gyroscale fluctuations, we assume the fluctuations are Alfv\'enic and write
\begin{equation}
\delta v = \frac{\sigma v_A \delta B}{B}
\label{eq:delta_v}
\end{equation}
where $\sigma = 1.19$ is a dimensionless constant used by \citet{Chandran_2010} for a spectrum of randomly phased kinetic Alfv{\'e}n waves with $k_\perp \rho_\perp \sim 1$ and the Alfv\'en velocity is given by
\begin{equation}
     v_A = \frac{B}{\sqrt{\mu_0 (n_p + 4 n_\alpha) m_p}}.
\end{equation}
The magnetic field fluctuations at the proton gyroscale $\delta B$ are defined as
\begin{equation}
    \delta B = \Bigg[\frac{\pi}{C_0(n_{s12})}  \int_{e^{-0.5 f_p}}^{e^{0.5 f_p}} P_B(f) df \Bigg]^{1/2} \label{eq:delta_B}
\end{equation}
where $f_p = v_{sw} \sin{\theta}/\rho_p$ is the convected gyrofrequency, $P_B(f)$ is the magnetic field power spectra,
\begin{eqnarray}
C_0(n_{s12}) = \frac{C_0(n_{s1}) \int_{e^{-0.5 f_p}}^{f_b} P_B(f) df + C_0(n_{s2})\int_{f_b}^{e^{0.5 f_p}} P_B(f) df}{\int_{e^{-0.5 f_p}}^{f_b} P_B(f) df + \int_{f_b}^{e^{0.5 f_p}} P_B(f) df} \\
C_0(n_{s}) = \int_0^{\pi/2} \cos{\phi}^{n_{s} - 1} d\phi = \frac{\pi^{0.5} \Gamma\big[{\frac{n_{s}}{2}}\big]}{2 \Gamma\big[{\frac{n_{s}+1}{2}}\big]}
\end{eqnarray}
are geometric terms described in \cite{Bourouaine_2013} and \cite{Vech_2017}; $f_b$ is the break frequency between the inertial and dissipation ranges.\footnote{Equation \ref{eq:delta_B} is similar to Equation 2 from \citet{Vech_2017}, but the expression in that article contains a typographical error. This error does not affect any results presented in that work (D. Vech, personal communication).}

To evaluate Equation \ref{eq:delta_B}, we process the E2 magnetic field power spectra $P_B(f)$ in the same way as done by \citet{Bourouaine_2013}, applying the algorithm illustrated in Figure \ref{fig:Spectrum}(b). First, we calculate a standard trace power spectrum as a sum of fast Fourier transforms of each of the three magnetic field components (blue solid line). The instrument sampling rate is not constant, but rather divided into regimes with timesteps $\Delta t \approx n 0.25$s, with $n$ a positive integer. For this analysis we make use only of periods where the resolution was $\Delta t \approx 0.25$s or $\approx 0.5$s, for a Nyquist frequency to 2 or 1 Hz, respectively. The peak at $f=1$ Hz due to the spacecraft spin is removed by a notch filter. The frequency range is then divided into logarithmically spaced regions and averaged within each of these regions (orange solid line).

From the processed spectrum, we look for the break point frequency $f_b$ in the logarithmic slope that is characteristic of the transition between magnetohydrodynamic (MHD) and ion scales in a similar fashion as was done by \cite{Vech_2017} (see Figure 1 (a) of that article). Throughout each spectrum, we perform a series of linear fits for logarithmic windows that contain $10\%$ of the data points. We define the starting frequency of the fit with the largest slope, $n_{s2}$, to be $f_b$.  We then perform a new linear fit for $f<f_b$ to find logarithmic slope $n_{s1}$ for the inertial range. These fits provide the power spectral density $B^2 / \nu$ at 1Hz, $P_{c1}$ and $P_{c2}$, for the portions of the spectra below and above $f_b$, respectively. Due to the possibility that $f_b$ is larger than the Nyquist frequency or is covered by instrument noise, we model $P_B$ as the piecewise function
\begin{eqnarray}
 \nonumber P_B(f) = 10^{P_{c1}} f^{-n_{s1}}, f<f_b \\
P_B(f) = 10^{P_{c2}} f^{-n_{s2}}, f>f_b
\end{eqnarray}
shown as a green dashed line on Figure \ref{fig:Spectrum}(b).
We reject any intervals where $n_{s2} > 1.7$ to ensure we resolve the dissipation scale spectra. We emphasize that the described procedure is not compromised by removing the 1 Hz peak as $f_b$ values, not plotted here, remain in the interval 0.2-0.4 Hz throughout the entire mission, as was elaborated in detail by \citet{Bourouaine_2012_ApJ}.

Given a value for $\delta B$ extracted from $P_B(f)$ (Equation \ref{eq:delta_B}), we calculate $\epsilon$ and $Q_\perp$
using Equations \ref{eq:Epsilon} and \ref{eq:Heating_Rate}.
From total of 259801 10-minute intervals, we have 82881 with $\Delta t \leq 0.5s$, 24615 of which satisfy the above described criteria. Out of these, 14954 intervals with  $\beta_{\parallel} < 0.3$ are used in this study as Equation \ref{eq:Heating_Rate} is derived in the low-$\beta$ limit (see Section \ref{sec:Intro}). In Figure \ref{fig:Spectrum}(a) we show the number of intervals used as a function of radial distance, using 20 linear bins $0.035$ au wide. Notably, the percentage of usable measurements significantly decreases for $r>0.6AU$. This is due to instrument related issues, as described in the following Section.

\section{Properties and limitations of the E2 magnetometer}
\label{sec:Instrumentation}

\begin{figure*}
\gridline{\fig{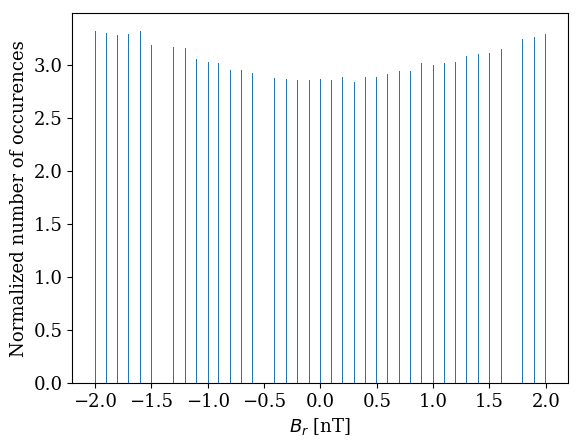}{0.49\textwidth}{(a)}
          \fig{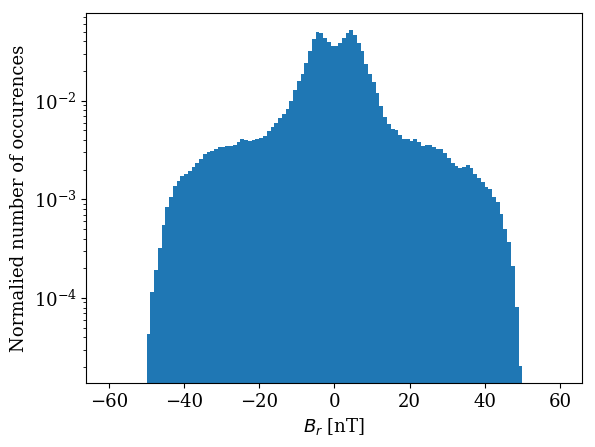}{0.49\textwidth}{(b)}
          }
\gridline{\fig{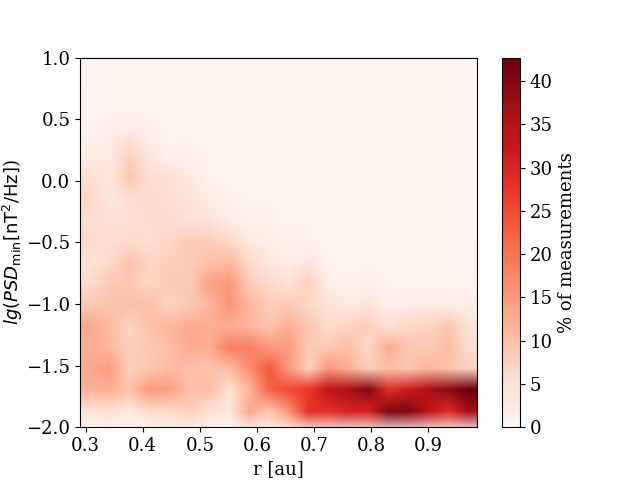}{0.49\textwidth}{(c)}
          \fig{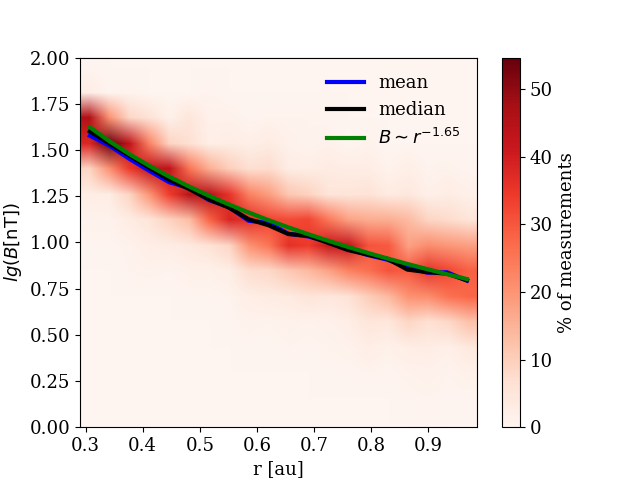}{0.49\textwidth}{(d)}
          }
\gridline{\fig{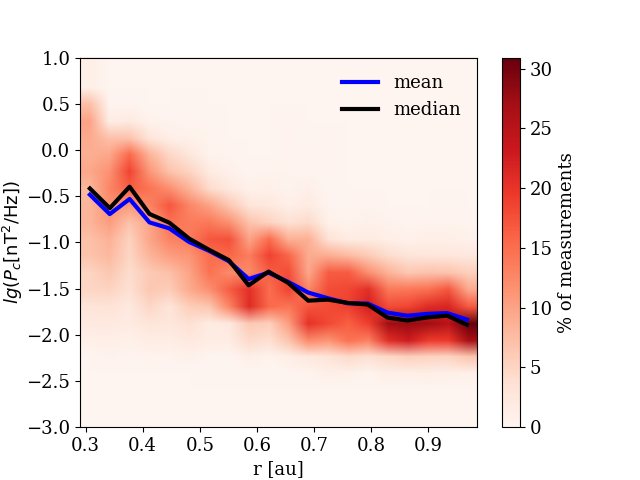}{0.49\textwidth}{(e)}
          \fig{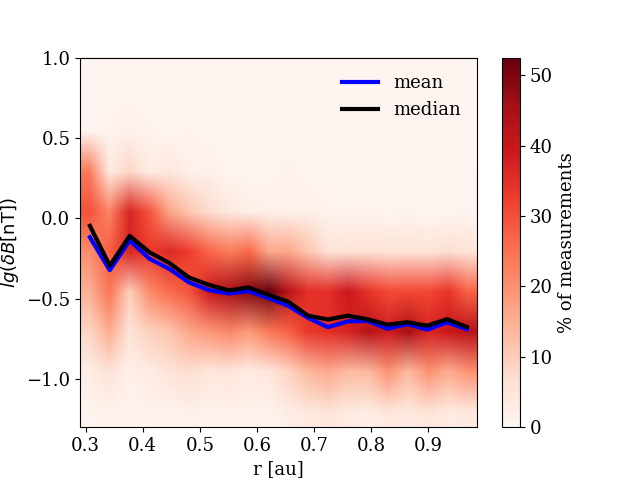}{0.49\textwidth}{(f)}
          }
    \caption{Histogram of the magnetic field radial component focused around zero (a) and over the full dynamic range (b). Note that no measurements record $B_r > 50$ nT. Therefore, the distributions of PSD minimum (c), $|B|$(d), $P_{c}$(e) ($P_{c1}$ if $f_b>1$ Hz or $P_{c2}$ if $f_b<1$ Hz) and $\delta B$ (f) show dips at $r \approx 0.3$. The flattening of the results at $r>0.65$ au is due to limited E2 sensitivity. Here, lg stands for base 10 logarithm.}
    \label{fig:B_histograms}
\end{figure*}

As the Helios instrument documentation doesn't provide a noise level for E2 magnetometers\footnote{The exact determination of the E2 noise level is an ongoing task of the Helios data archive team \citep{Cologne_2017}} it was necessary to extract it from the data. The available E2 data set provides measurements with accuracy of 0.1 nT for each of the magnetic field components. To check for a possible noise floor, we produced histograms of the radial component on Figure \ref{fig:B_histograms}, with bin sizes of 0.01 nT (a) and 0.1 nT (b), using total of $1.7 \times 10^8$ measurements with no averaging over time intervals. It is notable from panel (a) that there is no concentration point at $B_r = 0$. As the instrument sensitivity, extracted from the data, is $B_{s} = 0.1$ nT (the radial components can only have values $B_r = n B_s$, where $n$ is an integer), we conclude that the E2 noise level is less or equal to this value. The sensitivity $B_s$ has major influence on the data processing, as it sets the lowest possible level of magnetic field power spectral density (PSD), which is independent of frequency, to $B_s^2$ (panel (c)). Histograms for the other two components, not shown, behave in a similar fashion. Another important point is that no measurements record $B_r > 50$ nT. In the instrument team reports, it is noted that saturation might occur at this level, which is confirmed on panel (d).

These instrumental limitations have important consequences on the reliability of the measurements. From panel (d), we see the magnetic field value trend of $B \sim r^{-1.65}$ that has previously been reported beyond $r \approx 0.35$ au (see e.g. \citep{Hellinger_2013}). Below this distance, the magnetometer saturates and any increase in the measured field intensity comes only from tangential and normal components. This has a  major influence on derived values of $\delta B$ and, consequently, $\delta v$ and $\epsilon$, leading to significant underestimation, as shown on panel (f). 

On the other hand, at radial distances $r>0.65$ au, it is notable that, even though measured magnetic field values follow the expected radial trend, $\delta B$ results reach a constant value due to the E2 sensitivity floor. In order to verify this statement, we analyzed minimum of the power spectra, as well as behavior of the 1 Hz cutoff data, panels (c) and (e) respectively, where a floor is found starting at $r \approx 0.65$ au as well as a local minimum at $r \approx 0.3$ au are visible. We have much less confidence in results that come from these two regions, and consider them separately from the region between $0.35$ and $0.65$ au. Note that the percentage of intervals fitting our criteria significantly decreases for larger radial distances (Figure \ref{fig:Spectrum} (a)) even though the total number of results remains similar due to the spacecraft spending more time in this region.

\section{Results of Helios statistical study}
\label{sec:Results}

\begin{figure*}
\gridline{\fig{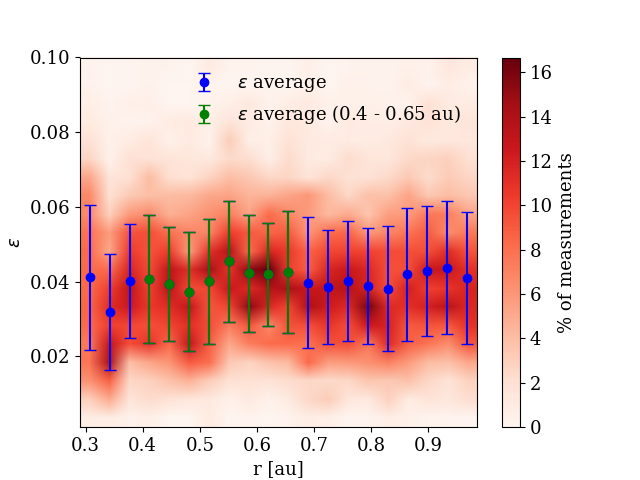}{0.49\textwidth}{(a)}
          \fig{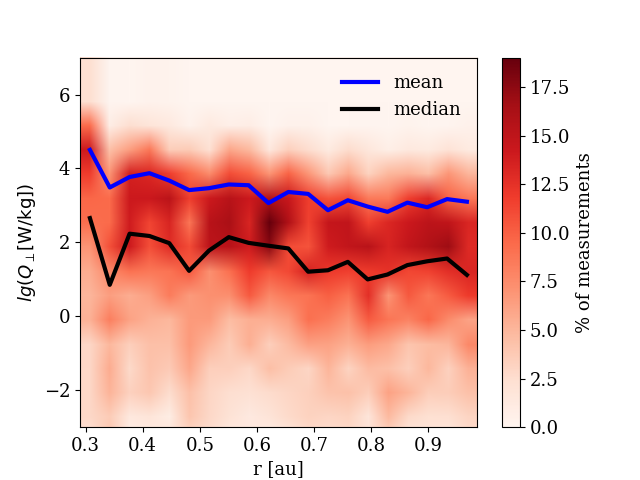}{0.49\textwidth}{(b)}
          }
\gridline{\fig{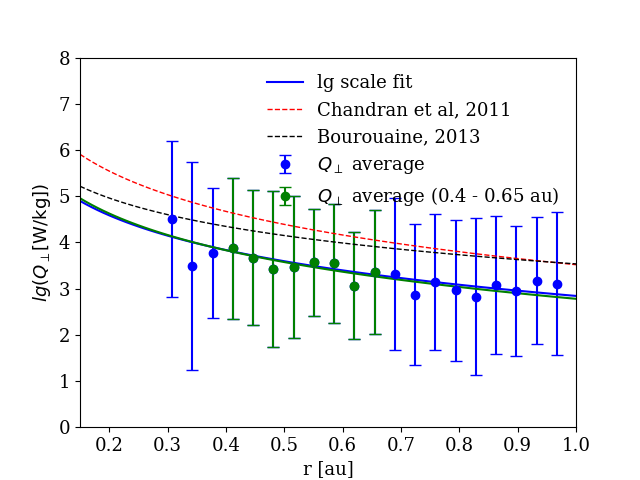}{0.49\textwidth}{(c)}
          \fig{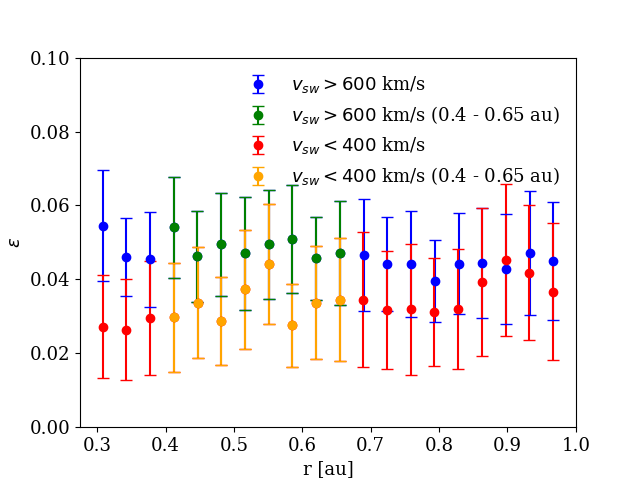}{0.49\textwidth}{(d)}
          }
\gridline{\fig{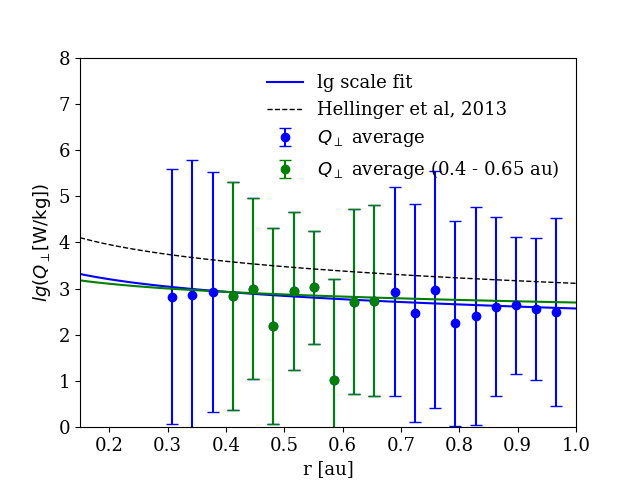}{0.49\textwidth}{(e)}
          \fig{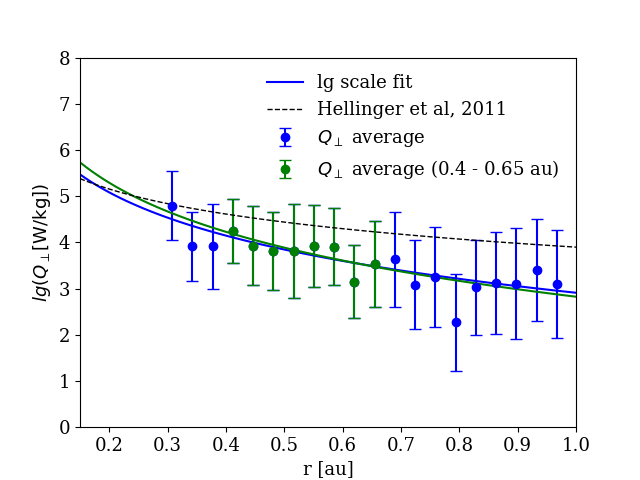}{0.49\textwidth}{(f)}
          }
    \caption{Histograms for each of the radial distance bins for $\epsilon$ and $Q_\perp$, along their average values and standard deviations used as uncertainties (a-d). Radial distances not affected by magnetometer noise levels or saturation, as described in Section \ref{sec:Instrumentation}, are given in green and orange dots. Panel (d) shows values of $\epsilon$ satisfying $v_{sw}<400$km/s or $v_{sw}>600$km/s, and the associated $Q_\perp$ compared with their respective empirical models are given in panels (e) and (f)}
    \label{fig:Results}
\end{figure*}

The results of our statistical study are shown on Figure \ref{fig:Results}. Mean values of $\epsilon$, shown on panel (a), are constant throughout the range of $r = 0.4 - 0.65$ au, where the Helios magnetometer observations are reliable. Using the derived $\epsilon$, we calculate $Q_\perp$ (panel (b)), using $c_1=0.75$ and $c_2=0.34$, values extracted from the test particle simulations by \citet{Chandran_2010}. As discussed in Section \ref{sec:Discussion}, while the value of these parameters can have significant influence on the amplitude of $Q_\perp$, they are expected to remain constant as a function of radial distance, and thus will not affect radial trends in $Q_\perp$.

The average value of $Q_\perp$ decreases with radial distance (red line), and can be logarithmically fit as $Q_\perp \sim r^{-2.5\pm0.4}$. We compare our results with two previous models of solar wind heating. A simple model used by \citet{Bourouaine_2013} (see Equation 6 of that article), when applied to our data set, reports a weaker radial trend of $Q_\perp \sim r^{-2.08}$ (black dashed line on panel (c)). On the other hand, a two-fluid theoretical model by \citet{Chandran_2011} that includes the effects of SH, as well as proton and electron Landau damping, reports $Q_\perp \sim r^{-2.9}$ as the perpendicular heating rate (green dashed line). It also predicts that the relative importance of SH compared to the electron heating should decrease beyond $r > 0.3 AU$. We note that the authors of this model state that its primary purpose was to predict conditions closer to the Sun, and that it may become increasingly inaccurate for larger radial distances due to violation of assumptions on the energy cascade times and relation between Elsass{\" e}r variables. 


The average measured amplitude of the heating, for $c_1=0.75$ and $c_2=0.34$ and considering only radial distances with reliable measurements (green dots on Figure \ref{fig:Results} (c)), is found to be $\sim 10^3 W/kg$ at 1 au, increasing up to $4.8 \times 10^4 W/kg$ at 0.15 au, with the radial trend described above. Comparing it to the results from \citep{Chandran_2011} and \citep{Bourouaine_2013}, we note that the contribution of SH from low $\beta$ solar wind streams is an order of magnitude below the total heating values predicted by these authors, increasing to $20-40\%$ at 0.15 au. However, these contributions are strongly dependent on the choice of the model constants (see Section \ref{sec:Discussion}).

In order to compare our results with available models obtained from observations, we use empirical total heating rates $Q(r)$ derived by \citet{Hellinger_2011} for $v_{sw} > 600$ km/s and \citet{Hellinger_2013} for $v_{sw} < 400$ km/s. We recalculate $\epsilon$ and $Q_\perp$ using only intervals that meet these bulk velocity criteria. For the slow wind, where 4325 out of 14954 intervals are used, the estimated SH rate is decreased compared to the total empirical heating rate ($Q_\perp \sim r^{-0.6 \pm 0.4}$ against $Q \sim r^{-1.2}$), as shown on panel (e). However, results for the fast solar wind, using 4225 out of 14954 intervals, are fundamentally different. Average values of $\epsilon$ are systematically higher in this case (panel (d)), which increases estimated $Q_\perp$ for the fast wind by an order of magnitude, as shown on panel (f), demonstrating that the majority of total heating, as well as the contribution from SH, happens in the fast solar wind streams. This amplitude increase is in agreement with the empirical model, but the radial trend of our derived stochastic heating rate in the fast wind is much steeper, $Q_\perp \sim r^{-3.1 \pm 0.5}$ compared to the empirical $Q \sim r^{-1.8}$. Therefore, the increasingly important role of SH at closer radial distances to the Sun is driven by intervals of fast solar wind. Additionally, since the Helios aphelion (with operational instruments) is at $r \approx 0.98$ au, this distance is the limit of the observational models compared with our data. The heating rates extracted here also compare well to the results of \citet{MacBride_2008_ApJ}, who followed the formalism developed by \citet{Politano_1998_PhRvE,Politano_1998_GeoRL} to account for the total dissipation rate of the turbulent energy in the inertial range. By studying 7 years of ACE data at L1, they found similar dissipation rates to those theoretically predicted \citet{Chandran_2011,Bourouaine_2013} and extrapolated from observations at smaller radial distances \citet{Hellinger_2011,Hellinger_2013}.

\begin{figure*}
\gridline{\fig{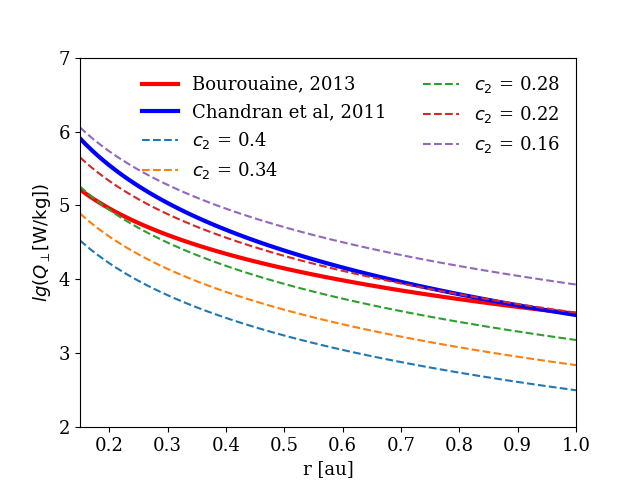}{0.49\textwidth}{(a)}
          \fig{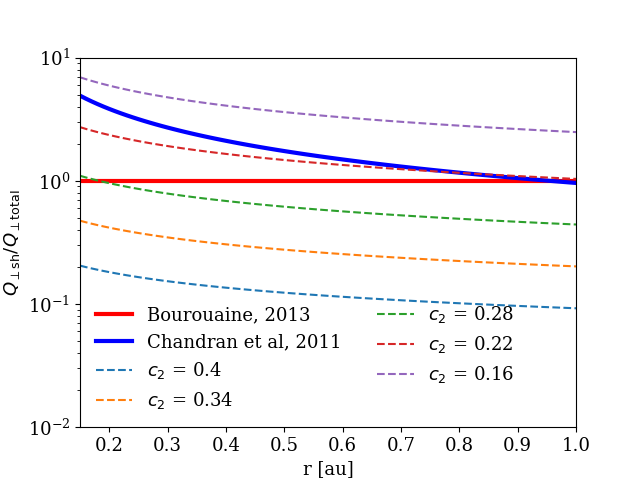}{0.49\textwidth}{(b)}
          }
    \caption{Dependence of the total SH rate on the assumed $c_2$ parameter. The orange dashed line on panel (a) is identical to the green solid line on Figure \ref{fig:Results} (c). Panel (b) illustrates the ratio $Q_{\perp \rm{SH}}$ and $Q_{\perp \rm{total}}$ from \cite{Bourouaine_2013} and \citep{Chandran_2011}.}
    \label{fig:c2}
\end{figure*}

Another major feature of our results is the spread of $Q_\perp$ values throughout 10 orders of magnitude. The shape of 1D histograms of $lg(Q_\perp)$ (not shown) obeys Gaussian distribution with low power tails, while $Q_\perp$ distributions are highly skewed with large differences between mean and median, as shown on Figure \ref{fig:Results} (b), with only $15\%$ of the measured intervals having amplitudes larger than the average value. These high values determine the bulk SH characteristics and will be separately discussed in the following Section.




\section{Discussion}
\label{sec:Discussion}

Following the method and results presented in Sections \ref{sec:Method} and \ref{sec:Results} there are several features and limitations of that demand further discussion. First, we only consider low $\beta$ plasma streams, which are, on average, expected to contribute less to $Q_\perp$ then cases with higher $\beta$, according to recent theoretical results \citep{Hoppock_2019}. The effect of SH in plasma with $\beta \gtrsim 1$ will be investigated in future work \citep{Martinovic_2019_ApJL}.

Variations in the three dimensionless parameters used in this model, $\sigma$, $c_1$ and $c_2$, may induce a systematic error in our results.
We assumed that only Alfv{\' e}nic fluctuations are present, and therefore asserted a relation between magnetic and velocity fluctuations in Equation \ref{eq:delta_v}, setting a constant value of $\sigma =1.19$. This parameter is expected to be lower for the case of non-Alfv{\' e}nic fluctuations (see e.g. \citep{Barnes_1974_JGR,Roberts_1987_JGR}), potentially becoming less than unity. However, by comparing panels (e) and (f) on Figure \ref{fig:Results}, it is notable that the highest dissipation rates are observed in the fast solar wind, where protons exhibit strong temperature anisotropy and highly Alfv{\' e}nic properties \citep{Stansby_2019_MNRAS}. We therefore argue that usage of the constant $\sigma$ is still justified for obtaining average $Q_\perp$. This enables us to predict that SH has an increasingly important role in the total solar wind heating rate at $r<0.3$ au, even though combined results shown on panel (c), along with very large uncertainties, indicate that its relative importance remains approximately the same for $r = 0.35 - 0.65$ au. Additional concern arises from some recent hybrid MHD simulation results \citep{Franci_2015_ApJ}, showing that magnetic and velocity fluctuations tend to decouple at the proton inertial length. This issue will be investigated through variations of $\sigma$ in future work.

The values of $c_1=0.75$ and $c_2=0.34$ are chosen using results from test-particle simulations. Values for these parameters have also been extracted from 
RMHD simulations \citep{Xia_2013}, with results summarized in Table 1 of that article and in Figure 5 of \citet{Bourouaine_2013}. These works argue that realistic values for $c_2$ can be as low as 0.15, similar to the value of $c_2=0.17$ assumed by \citet{Chandran_2011}. We examine this range of values below, with the understanding that additional refinements to values of $c_1$ and $c_2$ may become available from more realistic Vlasov simulations.

As $c_2$ participates in the exponential term in Equation \ref{eq:Heating_Rate}, its variation has the largest impact on $Q_\perp$. 
We examine its effect on $Q_\perp$ by recalculating Figure \ref{fig:Results} (c) using $c_2=0.16-0.4$ separated by increments of 0.06; this range covers parameter values considered by previous authors. As shown in Figure \ref{fig:c2}, varying $c_2$ by a factor of 2.5 increases $Q_\perp$ by one and a half orders of magnitude. 
We note that as $\epsilon$ linearly scales with $\sigma$, variations of $c_2$ can be used to examine variations in $\sigma$. 
Test particle simulations by \citet{Chandran_2010} have demonstrated that $c_1$ and $c_2$ are fairly insensitive to plasma parameters (see Figure 4 of that article) so we do not expect significant radial variation of those parameters. Note that for the case of $c_2=0.22$, $Q\perp$ is consistent with the heating rate from the \citep{Bourouaine_2013} model at large radial distances, but then exceeds the model by almost a factor of 2 at smaller distances. Therefore, if $Q_\perp$ is not to exceed $Q_{\rm{total}}$ at small radial distances, it is expected contribute relatively less and less at larger radial distances for values of $c_2>0.22$. On the other hand, comparison with the \citep{Chandran_2011} model, which includes both proton heating due to Landau damping and SH and electron heating, has a steeper radial dependence than either the \citep{Bourouaine_2013} model or our observations, matching our model for $c_2 = 0.22$ but exceeding it by a factor of 3 at $r = 0.2$ au. Comparison with this model states that values $c_2<0.22$, with radial trends calculated in Section \ref{sec:Results}, can be realistic within the range of uncertainties of our results, shown on Figure \ref{fig:Results} (c).

Even though is was noted above that the results for $r<0.35 AU$ are not reliable due to saturation in magnetic field measurements, we will still comment on the measured $\epsilon$ values for the radial bin $r<0.3AU$. The E2 saturation is expected to cause an underestimation of the measured $\delta B$, $\delta v$ and $\epsilon$, as seen in the bin at $r \sim 0.3$. However, the measured $\epsilon$ is one of the largest in Figure \ref{fig:Results} (a). This hints that SH might be the dominant process close to the Sun, and potentially the process active in the zone of preferential ion heating predicted by \citet{Kasper_2017} to extend tens of solar radii from the Sun's surface. Given the instrumental limitations described earlier in this work, this conclusion should be considered tentative and must verified by observations by future missions. Parker Solar Probe \citep{Fox_2016} will offer the first opportunity to provide such verification.

A last effect should be examined in future work. 
The applicability of stochastic heating in highly intermittant turbulence, where the fluctuating fields are larger than their rms values, potentially leading to increased heating \citep{Dmitruk_2004,Chandran_2010}, is an open question.
Recent results \citep{Mallet_2019} indicate that the effects of intermittency can be incorporated into models for nonlinear heating rates and play an important role in enhancing the solar wind heating due to nonlinear mechanisms such as SH.
This model will be confronted with solar wind observations, and are the focus of future work.

\acknowledgments

M.M.M. was financially supported by the Ministry of Education, Science and Technological Development of Republic of Serbia through financing the project ON176002.
K.G.K was supported by NASA grant NNX16AM23G. The authors would like to thank Chadi Salem and Daniel Vech for clarifying conversations. The data used for this project were sourced from NASA CDAweb \citep{CDAWeb_2018b} and the SSL Helios repository \citep{Cologne_2017_E2}. 

\bibliographystyle{aasjournal}
\bibliography{Latex_Refs}

\end{document}